\documentstyle[pramana,epsfig,floats]{ias}
\def\be{\begin{eqnarray}}
\def\ee{\end{eqnarray}}
\def\beq{\begin{equation}}
\def\eeq{\end{equation}}
\def\ba{\begin{array}}
\def\ea{\end{array}}
\def\hc{{\rm H.c.}}
\def\bec{\begin{center}}
\def\ec{\end{center}}

\begin{document}
\mark{{Partially composite 2-Higgs doublet model }{Dong-Won Jung}}
\title{Partially composite 2-Higgs doublet model}

\author{Dong-Won Jung}
\address{School of Physics, KIAS Seoul 130-722, Korea}

\abstract{In the extra dimensional scenarios with gauge fields in
the bulk, the Kaluza-Klein (KK) gauge  bosons can induce
Nambu-Jona-Lasinio (NJL) type attractive four-fermion
interactions, which can break electroweak symmetry dynamically
with accompanying composite Higgs fields. We consider a
possibility that electroweak symmetry breaking (EWSB) is triggered
by both a fundamental Higgs and a composite Higgs arising in a
dynamical symmetry breaking mechanism induced by a new strong
dynamics. The resulting Higgs sector is a partially composite
two-Higgs doublet model with specific boundary conditions on the
coupling and mass parameters originating at a compositeness scale
$\Lambda$. The phenomenology of this model is discussed including
the collider phenomenology at LHC and ILC. }

\maketitle

\section{Introduction}
 Although the standard model (SM) has been successful to describe the physics
up to $E \sim 200$ GeV, the origin of the electroweak symmetry
breaking(EWSB) still remains as a mystery. Among the various
models for the EWSB, the extra dimensional scenarios have
 many interesting characteristics. Under this
extra dimensional setup, the EWSB can occur in various ways. It
can be accomplished by fundamental Higss bosons, and/or some
dynamical way by bulk QCD, and/or nontrivial boundary conditions,
etc.. It is an interesting observation that in the extra
dimensional scenarios the EWSB can be achieved by both fundamental
Higgs boson and dynamical fermion condensates. With this
philosophy in mind, we study an extension of Bardeen-Hill-Lindner
(BHL) scenario \cite{BHL}. The BHL scenario is particularly
interesting since the heavy top mass is intimately related with a
new strong dynamics that condenses the $t\bar{t}$ bilinear, and
breaks the EW symmetry. The original version of BHL scenario
predicts that the top mass should be larger than $\sim 200$ GeV,
which is no longer viable. The extension of BHL with two composite
Higgs doublets has a similar shortcoming \cite{luty}. In our model
\cite{Chung:2005yz}, a fundamental scalar particle is introduced
in addition to the $t\bar{t}$ condensate, and we successfully fit
the top and bottom quark masses simultaneously. Our model predicts
$\tan \beta$ between the 0.45 and 1. In addition, our model can be
tested in the future collider experiments.

\section{Model Description}

We introduce a strong dynamics to the standard model at some high
scale  $\Lambda$, which is  effectively described by the NJL type
four-fermion interaction term.

\be {\cal L} = {\cal L}_{\rm SM} +
      G ( \overline{\psi}_{L} t_{R} )( \overline{t}_{R} \psi_{L} ),
\ee where \be {\cal L}_{\rm SM} = {\cal L}_{\rm gauge} + {\cal
L}_f + {\cal L}_{\phi}
            + (y_{t0}~\overline{\psi}_L t_R \tilde{\phi} + \hc)
            + (y_{b0}~\overline{\psi}_L b_R \phi + \hc)
\ee and $\tilde{\phi} \equiv i \sigma_y \phi^*$.
 Introducing an auxiliary scalar doublet field
$\Phi (x)$, we can rewrite the NJL term in Eq.(2.1) as \be {\cal
L} = {\cal L}_{\rm SM}
           + g_{t0} ( \overline{\psi}_{L} t_{R} \tilde \Phi + \hc )
           -M^2 \Phi^{\dagger} \Phi,
\ee where $G = g_{t0}^2 / M^2$ and $g_{t0}$ is a newly defined
Yukawa coupling. The mass scale $M$ will be generically of order
$\Lambda$. The new scalar field $\Phi$ describes the composite
scalar bosons that appear when the $\langle \bar{t} t \rangle$
develops nonvanishing VEV and breaks the electroweak symmetry.
Then we have one fundamental scalar field $\phi$ and one composite
scalar field $\Phi$, although $\Phi$ is not a dynamical field at
the scale $\Lambda$. Far below the scale $\Lambda$, the $\Phi$
field will develop the kinetic term due to quantum corrections and
become dynamical. The resulting low energy effective field theory
will be two-Higgs doublet model. The large FCNC problem can be
avoided by assigning a $Z_2$ discrete symmetry \cite{weinberg}
\begin{eqnarray*}
( \Phi,~\psi_L,~U_R ) & \to & + ( \Phi,~\psi_L,~U_R ) , ~ ( \phi,
D_R ) \to  - ( \phi, D_R ) .
\end{eqnarray*}
Then the Yukawa term $(y_{t0}~\overline{\psi}_L t_R \phi + \hc)$
is forbidden and only the $y_{b0}$ coupling term remains. From now
on, we will rename $y_{b0}$ as $g_{b0}$. In consequence, our model
becomes the Type-II two-Higgs doublet model.
 We write the effective
lagrangian far below $\Lambda$ as \be {\cal L} &=& {\cal L}_{\rm
gauge} + {\cal L}_f
               + (D_\mu \phi)^\dagger(D^\mu \phi)
               + (D_\mu \Phi)^\dagger(D^\mu \Phi)
               \nonumber \\
               &&~~~~~~
               + (g_b \overline{\psi}_{L}b_R \phi + {\rm H.c})
               + (g_t \overline{\psi}_{L}t_R \tilde \Phi + {\rm H.c})
               - V(\phi, \Phi),
\ee where the most general Higgs potential is given by \be V(\phi,
\Phi) &=& \mu_1^2\phi^\dagger\phi + \mu_2^2\Phi^\dagger \Phi
              + ( \mu_{12}^2  \phi^\dagger \Phi + \hc)
         + \frac{1}{2}\lambda_1(\phi^\dagger\phi)^2
              + \frac{1}{2}\lambda_2(\Phi^\dagger \Phi)^2
\nonumber \\
         &&~+ \lambda_3(\phi^\dagger\phi)(\Phi^\dagger \Phi)
              + \lambda_4|\phi^\dagger \Phi|^2
              + \frac{1}{2} [ \lambda_5 (\phi^\dagger \Phi)^2 + \hc ].
\ee In the scalar potential, we have introduced a dimension-two
$\mu_{12}^2$ term that breaks the $Z_2$ discrete symmetry softly
in order to generate the nonzero mass for the CP-odd Higgs boson.
This $\mu_{12}^2$ parameter will be traded with the $m_A^2$. The
renormalized at low energy is given by \be {\cal L}_{\rm ren} &=&
Z_\phi(D_\mu\phi)^\dagger(D^\mu \phi)
          + Z_\Phi (D_\mu \Phi)^\dagger(D^\mu \Phi)
          - V(\sqrt{Z_\phi}\phi, \sqrt{Z_\Phi}\Phi)
\nonumber \\
  &&~~~+ \sqrt{Z_\Phi}g_t(\overline{\psi}_L t_R \tilde \Phi + {\rm h.c})
          + \sqrt{Z_\phi}g_b(\overline{\psi}_L b_R \phi + {\rm h.c}),
\ee and matching the lagrangian with Eq. (2.3) at the
compositeness scale $\Lambda$, we obtain the matching condition
\be && \sqrt{Z_\phi}\rightarrow 1,~~~~~ \sqrt{Z_\Phi} \rightarrow
0, ~~~~~ Z_\phi \mu_1^2 \rightarrow m_{0}^2,~~~~~ Z_\Phi \mu_2^2
\rightarrow M^2,
\\
&& Z_\phi \lambda_1 \rightarrow \lambda_{10},~~~~~
Z_\Phi^2\lambda_2 \rightarrow 0,~~~~~Z_\phi Z_\Phi
\lambda_{i=3,4,5} \rightarrow 0, \nonumber \ee as the scale $\mu
\rightarrow \Lambda$. Note that our model has non-vanishing
$\sqrt{Z_\phi}$ and $Z_\phi \lambda_1$ unlike the Luty's model, we
can fit both the bottom and top quark masses.

\section{Particle Spectrum}

%\begin{figure}[t]
%\begin{center}
%\hbox
%to\textwidth{\hss\epsfig{file=fig3.eps,width=6cm,height=4cm}\hss}
%\vspace{0.2cm} \caption{ Allowed values of the quartic coupling
%$\lambda_{10}$ at $\Lambda$ with respect to the compositeness
%scale $\Lambda$. }
%\end{center}
%\end{figure}

 Our model is defined in terms of three parameters: Higgs self
coupling $\lambda_{10}$, the compositeness scale $\Lambda$ (where
$\lambda_{10}$ and the NJL interaction are specified), and the
CP-odd Higgs boson mass $m_A$. Using the field redefinition \be
\phi \rightarrow Z_\phi^{-1/2}\phi,~~~~~ \Phi \rightarrow
Z_\Phi^{-1/2}\Phi, \ee we rewrite the matching condition given in
Eq. (2.7) as \be && g_b \rightarrow g_{b0}, ~~~~~ g_t \rightarrow
\infty, ~~~~~ \lambda_1/g_b^4 \rightarrow \lambda_{10}/g_{b0}^4,
~~~~~ \lambda_{2,3,4,5} \rightarrow 0, \ee for the rescaled
couplings. These conditions are the boundary conditions for the RG
equations.
 We will use the one-loop RG equations given in Ref. \cite{hillleungrao}.
After numerical analysis, we find that the allowed parameter space
is rather restricted as the increase of $\Lambda$. Since the $m_A$
is a free parameter in our model, other Higgs boson masses can be
calculated in various parameter region. Again, we can get rather
narrow band for the Higgs masses as a function of $m_A$. For
$\Lambda \sim 10^{15}$ GeV, $m_h$ larger than 250 GeV is
predicted. Also the charged Higss can be lighter than $h$, and
could be the first signal of our model at future collders.
Generically charged Higgs boson is lighter than the CP-odd Higgs.

\section{Higgs Production at LHC and ILC}
\begin{figure}[t]
\begin{center}
\hbox to\textwidth{\hss
\epsfig{file=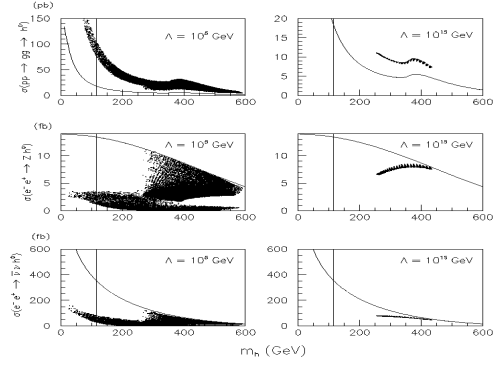,width=7cm,height=5.1cm} \hss} \vspace{0cm}
\caption{ Production  cross section of the neutral Higgs boson at
the LHC and ILC. $\sqrt{s} = 14$ TeV for the LHC and  $\sqrt{s} =
1$ TeV for the ILC are assumed. The solid curves denotes the SM
predictions. }
\end{center}
\end{figure}

The ratio of self-couplings to the SM values
${\lambda_{hhh}}/{\lambda^{\rm SM}_{hhh}},~
{\lambda_{hhhh}}/{\lambda^{\rm SM}_{hhhh}}$ are enhanced for the
small mass of the light Higgs boson $h$. In most parameter region,
our triple gauge coupling is negative and it can lead to many
interesting phenomenology. \cite{Chung:2005yz}

 The cross sections predicted in our model are modified from those in
the SM by factors \cite{djouadi}, \be \sigma (pp \rightarrow gg
\rightarrow h^0 ) &=& \left( \frac{\cos \alpha}{\sin \beta}
\right)^2 \sigma_{SM}(pp \rightarrow gg \rightarrow h^0 ),
\nonumber \\
 \sigma (e^+ e^- \to Z + h^0 / H^0) &=& \sin^2 ~/~ \cos^2
(\beta - \alpha) \sigma_{\rm SM} (e^+ e^- \to Z + h^0 / H^0),
\nonumber \\
\sigma (e^- e^+ \to \bar{\nu}_e \nu_e + h^0/H^0) &=& \sin^2 ~/~
\cos^2 (\beta - \alpha) \sigma_{\rm SM} (e^- e^+ \to \bar{\nu}_e
\nu_e + h^0/H^0). \ee In Fig. 1, we show the single Higgs boson
production cross sections at the LHC and the ILC for different
$\Lambda$ and $m_h$, etc.. For higher $\Lambda$, the allowed $m_h$
has a narrow region. Therefore the cross sections are almost
definitely determined for large $\Lambda$, and they will be a
strong signature of our model.

\section{Summary}

We considered an interesting possibility that the Higgs boson
produced at the future colliders is neither a fundamental scalar
nor a composite scalar, but a mixed state of them. This scenario
could be easily realized, if we embed the SM in a higher dimension
with bulk gauge interactions. We have constructed the simplest
model with the NJL type four-fermion interaction of top quarks as
the strong dynamics inspired by the BHL and study the
phenomenology of the two Higgs doublets model with the
compositeness condition as the low energy effective theory. The
resulting theory can accommodate the observed top mass, and give
specific predictions for neutral and charged Higgs masses at a
given value of $\Lambda$. The charged Higgs boson is always
lighter than the CP-odd Higgs neutral boson, although the mass
difference is very small. For $\Lambda \sim 10^{15}$ GeV, the
allowed parameter region is rather restricted, and we predict $m_h
> 250$ GeV. The cross sections for the Higgs boson production are
modified in our model, and it will be observed in the future
colliders, $i.e.$, LHC and ILC.

%%%%%%%%%%%%%%%%%% References
%%%%%%%%%%%%%%%%%%%%%%%%%%%%%%%%%%%%%%%%%%%%%%%%%%%%%%%%%%%%%%%%%%%%%%%
\def\PRD #1 #2 #3 {Phys. Rev. D {\bf#1},\ #2 (#3)}
\def\PRL #1 #2 #3 {Phys. Rev. Lett. {\bf#1},\ #2 (#3)}
\def\PLB #1 #2 #3 {Phys. Lett. B {\bf#1},\ #2 (#3)}
\def\NPB #1 #2 #3 {Nucl. Phys. {\bf B#1},\ #2 (#3)}
\def\ZPC #1 #2 #3 {Z. Phys. C {\bf#1},\ #2 (#3)}
\def\EPJ #1 #2 #3 {Euro. Phys. J. C {\bf#1},\ #2 (#3)}
\def\JHEP #1 #2 #3 {JHEP {\bf#1},\ (#2) #3}
\def\IJMP #1 #2 #3 {Int. J. Mod. Phys. A {\bf#1},\ #2 (#3)}
\def\MPL #1 #2 #3 {Mod. Phys. Lett. A {\bf#1},\ #2 (#3)}
\def\PTP #1 #2 #3 {Prog. Theor. Phys. {\bf#1},\ #2 (#3)}
\def\PR #1 #2 #3 {Phys. Rep. {\bf#1},\ #2 (#3)}
\def\RMP #1 #2 #3 {Rev. Mod. Phys. {\bf#1},\ #2 (#3)}
\def\PRold #1 #2 #3 {Phys. Rev. {\bf#1},\ #2 (#3)}
\def\IBID #1 #2 #3 {{\it ibid.} {\bf#1},\ #2 (#3)}
%%%%%%%%%%%%%%%%%%%%%%%%%%%%%%%%%%%%%%%%%%%%%%%%%%%%%%%%%%%%%%%%%%%%%%%


\begin{thebibliography}{999}


\bibitem{BHL} W. A. Bardeen, C. T. Hill and M. Lindner,
\PRD 41 1647 1990 ; C. T. Hill, Report No. FERMI-CONF-90/170-T.

\bibitem{luty} M. Luty, \PRD 41 2893 1990

%\cite{Chung:2005yz}
\bibitem{Chung:2005yz}
  B.~Chung, K.~Y.~Lee, D.~W.~Jung and P.~Ko,
  %``Partially composite two-Higgs doublet model,''
  JHEP {\bf 0605}, 010 (2006)
  [arXiv:hep-ph/0510075].
  %%CITATION = HEP-PH 0510075;%%

%\bibitem{extraD} N. Arkani-Hamed, S. Dimopoulos and G. R. Dvali,
%\PLB 429 263 1998 ; \IBID 436 257 1998 ; L. Randall and R.
%Sundrum, \PRL 83 3370 1999 ; \IBID 83 4690 1999 ; T. Appelquist,
%H.-C. Cheng and B. A. Dobrescu, \PRD 64 035002 2001 ; C. Csaki, C.
%Grojean, L. Pilo and J. Terning, \PRL 92 101802 2004 ; K. Agashe,
%A. Delgado, M. J. May and R. Sundrum, \JHEP 0308 050 2003 .

%\bibitem{hscalar} E. H. Simmons, \NPB 312 253 1989 ; V. Hemmige and E. H.
%Simmons, \PLB 518 72 2001 .

%\bibitem{bosontop} C. D. Carone and E. H. Simmons, \NPB 397 591 1993 ;
%C. D. Carone and H. Georgi, \PRD 49 1427 1994 ; K. G. Chetyrkin
%and O. V. Tarasov, \PLB 327 114 1994 ; C. D. Carone, E. H. Simmons
%and Y. Su, \PLB 344 287 1995 ; A. Aranda and C. D. Carone, \PLB
%488 351 2000 ; \IJMP 16S1C 899 2001 ; A. Kagan and S. Samuel, \PLB
%246 250 1990 ,
% \IBID 252 605 1990 , \IBID 270 37 1991 ,
%and \IBID 284 289 1992 ; \IJMP 7 1123 1992 .

%\bibitem{DSBlarge} H. Georgi, A. K. Grant and G. Hailu,
%\PLB 506 207 2001 ; \PRD 63 064027 2001 .

%\bibitem{DSBwarped} N. Rius and V. Sanz, \PRD 64 075006 2001 ;
%K. Oda and A. Weiler, \PLB 606 408 2005 .

%\bibitem{cskim} G. Cvetic, S. S. Hwang and C. S. Kim,
%\IJMP 14 769 1999 ; Acta Phys. Pol. B {\bf 28}, 2515 (1997).

%\bibitem{work1} Work in progress.

\bibitem{weinberg}
S.L. Glashow and S. Weinberg, Phys. Rev. D 15, 1958 (1977).

\bibitem{hillleungrao} C. T. Hill, C. N. Leung and S. Rao,
\NPB 262 517 1985 .

%\bibitem{pdg} S.~Eidelman {\it et al.}  [Particle Data Group],
%  %``Review of particle physics,''
%  Phys.\ Lett.\ B {\bf 592}, 1 (2004).
%  %%CITATION = PHLTA,B592,1;%%

%\bibitem{mu12}
%I.F. Ginzburg and M.V. Vychugin, Presented at the 16th
%International Workshop on High Energy Physics and Quantum Field
%Theory (QFTHEP 2001), Moscow, Russia, 6-12 Sep 2001. Published in
%the proceeding, "Moscow 2001, High energy physics and quantum
%field theory" p.64-76 hep-ph/0201117; I.F. Ginzburg, M. Krawczyk
%and P. Osland Presented at International Workshop on Linear
%Colliders (LCWS 2002), Jeju Island, Korea, 26-30 Aug 2002.
%Published in "Seogwipo 2002, Linear colliders" p. 90-94
%hep-ph/0211371.

%\bibitem{selfcoupling} M. Moretti, S. Moretti, F. Piccinini,
%R. Pittau and A. D. Polosa, \JHEP 0502 024 2005 ; S. Kanemura, Y.
%Okada and E. Senaha, \PLB 606 361 2005 .

%\bibitem{selfcoupling1}
%S.~Kanemura, Y.~Okada, E.~Senaha and C.~P.~Yuan,
%  %``Higgs coupling constants as a probe of new physics,''
%  Phys.\ Rev.\ D {\bf 70}, 115002 (2004)
%  [arXiv:hep-ph/0408364].
%  %%CITATION = HEP-PH 0408364;%%

%\bibitem{work2} Work in progress.

\bibitem{djouadi} See, for example, recent reviews and references therein:
A.~Djouadi,
  %``The anatomy of electro-weak symmetry breaking. I: The Higgs boson in the
  %standard model,''
  arXiv:hep-ph/0503172;
  %%CITATION = HEP-PH 0503172;%%
%A.~Djouadi,
  %``The anatomy of electro-weak symmetry breaking. II: The Higgs bosons in the
  %minimal supersymmetric model,''
  arXiv:hep-ph/0503173.
  %%CITATION = HEP-PH 0503173;%%

%\bibitem{c} F. Nesti and P. Dall'Aglio, \emph{Sample proceedings in
%                JHEP format}, SISSA 2001.
%\bibitem{fltf}  Maths Dahlgren, {\it Package {\tt floatflt}, distributed
%                with \LaTeXe{} 96/06/01}, 1994-1996.
%\bibitem{LC}    M. Goossens, F. Mittelbach, A. Samarin,
%                {\it The \LaTeX{} Companion}, Addison-Wesley 1994.
%\bibitem{TeXbook} D. E. Knuth, {\it The \TeX book}, Addison-Wesley 1986.


\end{thebibliography}
\end{document}